\documentclass[draft]{agujournal2019}
\usepackage{url} 
\usepackage{lineno}
\usepackage[inline]{trackchanges} 
\usepackage{soul}
\usepackage{natbib}
\usepackage{algorithm}
\usepackage[normalem]{ulem}

\usepackage{algpseudocode}
\usepackage{booktabs}

\usepackage[utf8]{inputenc}
\usepackage[T1]{fontenc}

\draftfalse

\journalname{Space Weather}

\begin{document}

%
%

\title{Post-processing Probabilistic Forecasts of the Solar Wind by Data Mining Similar Scenarios}


%
%




\authors{Daniel E. da Silva\affil{1,2}, Yash Parlikar\affil{1,3}, Shaela I. Jones\affil{1,4}, Charles N. Arge\affil{1}}


\affiliation{1}{Heliophysics Sciences Division, NASA Goddard Spaceflight Center, Greenbelt, MD, USA}
\affiliation{2}{Goddard Planetary Heliophysics Institute, University of Maryland, Baltimore County, MD, USA}
\affiliation{3}{Institute for Space Weather Sciences, New Jersey Institute of Technology, Newark, New Jersey, USA}
\affiliation{4}{Department of Physics, Catholic University of America, Washington, DC, USA}




\correspondingauthor{Daniel E. da Silva}{Daniel.E.daSilva@nasa.gov}
\renewcommand{\texthl}[1]{#1}



\begin{keypoints}   
\item A novel methodology extending analog ensembles is developed to produce probabilistic forecasts of the solar wind speed
\item The technique augments single-value prediction pipelines with recent \texthl{observational} information, and demonstrated/validated with ADAPT-WSA

\item  Implications of such empirical post-processing algorithms around single-value models in space weather forecasting is discussed 

\end{keypoints}

%
%

%
%


\begin{abstract}
The solar wind speed at Earth is one of the most important parameters regarding the effects of space weather on society.  Thus far, most approaches for predicting the solar wind speed produce a single-value time series without uncertainty, or utilize ensemble methods which require custom calibration development. In this study, a method is developed that produces calibrated probabilistic forecasts of the solar wind speed using skew normal distributions and a novel extension of analog ensembles. In our extension, the single-value predictions from a baseline model of the next $\Delta t$ days  are used along with $\Delta window$ hours of recent observations and single-value predictions to create a forecasting scenario vector that is compared against a historical database for outcomes. The baseline model used is the combined Air Force Data Assimilative Photospheric Flux Transport-Wang Sheeley Arge (ADAPT-WSA) model and the WSA point parcel simulation, but the method is directly applicable to other deterministic models including components such as Enlil or \texthl{the Heliospheric Upwind Extrapolation with time dependence model (HUXt)}. The approach works notably well on the benchmark of whether observations fall within the $p^{th}$ percentile $p\%$ of the time (for $p$ between 0 and 100). Falling back on the mean or median of the predicted distribution as a non-probabilistic prediction yields a direct improvement in \texthl{root-mean-square error (RMSE) over the original WSA point parcel simulation, and is shown to beat $\approx$ 1 solar rotation recurrence for 1-5 day ahead forecasts.}

\end{abstract}

\section*{Plain Language Summary}
The solar wind speed at earth is important for space weather, but usually isn't predicted with uncertainty. In this manuscript we create a method for adding forecasts with probability as a step after the prediction is made from another pipeline. The pipeline we base this demonstration off of is called Air Force Data Assimilative Photospheric Flux Transport-Wang Sheeley Arge (ADAPT-WSA) with the WSA point parcel simulation, but others can be used. What we come up with works very well when you test it on whether measurements of the solar wind fall within the pth percentile p\% of the time, for values of p between 0 and 100. Also, using the mean of the probabilistic forecast results in better errors than the baseline model.


%
%

\section{Introduction}
The solar wind is the single most important time-varying state that physically influences the onset of geomagnetic storms and geoeffective disturbances at Earth. The solar wind speed, $V$, plays a crucial role in the dayside compression of Earth's magnetosphere,  broadly compressing further with increasing dynamic pressure \citep{shue1998magnetopause}.  It has been established in previous research that  high-speed solar wind streams lead to the development of co-rotating interaction regions (CIRs) and variations in $B_z$ that cause magnetic reconnection at Earth, \texthl{the driver of geomagnetic storms} \citep{echer2008interplanetary,pulkkinen2007space}. Furthermore, enhancements in solar wind speed increase the energy transfer into Earth’s magnetosphere raising the potency of existing storms \citep{gonzalez1994geomagnetic,richardson2012solar,paulikas1979effects}.  \texthl{Specifically, the reconnection rate on the dayside is the rate at which southward magnetic flux is convected by the solar wind into the magnetosphere, given by $-V_x \times B_z$ where $V_x$ is the solar wind speed along the earth-to-sun line and and $B_z$ is the interplanetary magnetic field z-direction in Geocentric Solar Magnetospheric (GSM) coordinates} \citep{bothmer2007space}. \texthl{In addition, coronal mass ejections (CMEs) are another major driver of space weather due to their potential for large to extremely large flow speeds and southward $B_z$. While CMEs aren't the focus of this paper, CMEs do propagate through the solar wind, with their speed and arrival time depending on the background conditions. }For these reasons, forecasting the solar wind is essential for effective space weather prediction. 

Approaches to predicting the solar wind speed have largely centered on simulations that produce a single-value time-series \citep{arge2000improvement,arge2004stream,rotter2015real,sheeley2017origin,upendran2020solar}  or ensembles of single-value solutions intended to capture the variation due to unknown or unobservable quantities \citep{riley2013application,owens2017probabilistic,wang2018operational,vidal2023machine,edward2024adapting}. \texthl{Modeling approaches also include fully physics based model like Magnetohydrodynamic Algorithm outside a Sphere (MAS;} \citealp{riley2012corotating}) \texthl{and the Alfvén Wave Solar Atmosphere Model (AWSoM; }\citealp{huang2024solar}). The ability to attach uncertainty to a forecast-- that is, make a probabilistic forecast-- allows users of the forecast to create more sophisticated and calculated responses. This concept is well studied in weather forecasting \citep{ehrendorfer1997predicting,joslyn2012uncertainty,palmer2000predicting}. A large survey of the U.S. public, performed by \citet{morss2008communicating}, confirmed a significant majority of participants indeed appreciated uncertainty added to weather forecasts, and in fact preferred them to single-valued forecasts.

Methodologies for predicting time series with uncertainties are varied \citep{siddique2022survey}. Common approaches include Gaussian processes \citep{azari2024virtual},  or direct prediction of distribution parameters using neural networks \citep{camporeale2019generation}, and the previously mentioned ensemble modeling \citep{murray2018importance}.

\begin{figure}
    \centering
    \includegraphics[width=1\linewidth]{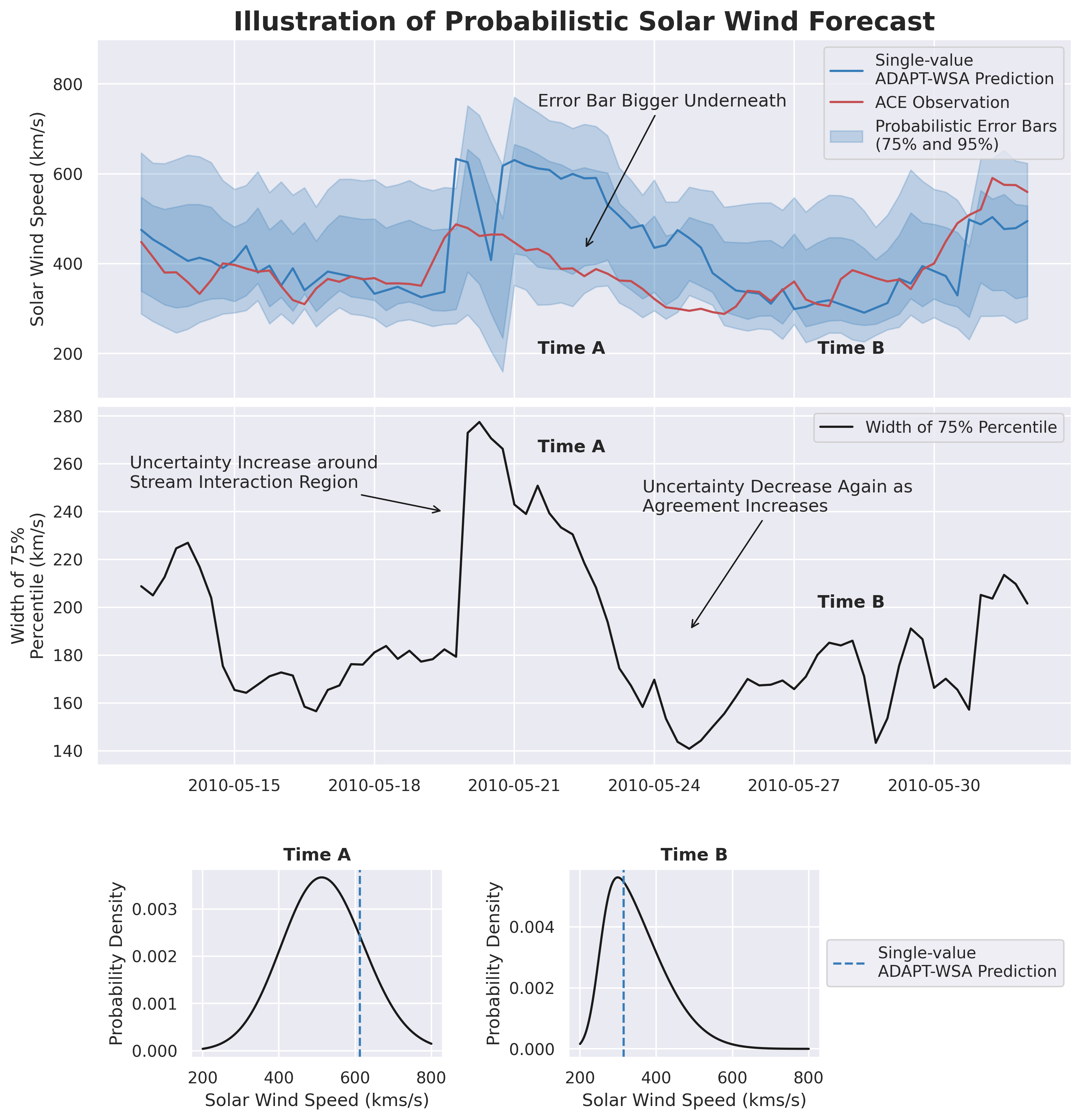}
    \caption{Illustration of probabilistic solar wind prediction and adaptive uncertainty using the method of this article, for predictions at Earth using a single ADAPT map (realization 0) and 3-day-ahead predictions. \texthl{Within this time interval is a stream interaction region predicted by ADAPT-WSA and observed by ACE around 2010-05-20}. The bottom two panels show the probabilistic predictions at \texthl{\textit{Time A} (2010-05-21, 12 UT) and \textit{Time B} (2010-05-27, 12 UT).}}
    \label{fig:illustration}
\end{figure}

In this article, we prepare calibrated probabilistic predictions of the solar wind using a methodology that augments an existing single-value model of the solar wind speed. Our methodology is developed using the combined Air Force Data Assimilative Photospheric Flux Transport-Wang Sheeley Arge (ADAPT-WSA) model.  The output of ADAPT, which are global magnetic field maps at the photosphere, is used as input to WSA. WSA models the coronal domain and is a semi-empirical model that predicts the speed and interplanetary magnetic field polarity of the solar wind \citep{arge2003improved,arge2004stream,mcgregor2008analysis,sheeley2017origin}. Included in WSA is a point-parcel simulation for propagation of the solar wind out into the heliosphere \citep{da2023ensemble}. 

While we use ADAPT-WSA, the methodology can be applied with minimal modification to other simulation systems provided a sufficient record of historical runs exists. More information on ADAPT-WSA can be found in Section \ref{sec:adaptwsa}.

This article's methodology for probabilistic predictions is based on defining a criterion for comparable forecasting scenarios, finding matching neighbors, and using those neighbors to fit a parametric skew normal distribution at each point in time. This is an extension to the method of analog ensembles, previously proposed in the literature for solar wind prediction in \citet{owens2017probabilistic,riley2017forecasting,simon2025analog}. 

In analog ensembles, the last $\Delta t$ days of the solar wind speed, or another solar wind plasma variable, are compared with a sliding window of $\Delta t$ days running back through a historical record. The closest $k$ neighbors are selected this way, with closeness defined as the Euclidean distance. The solar wind speeds following the sliding window are then used to form an ensemble of future predictions. 

The heritage of analog ensembles is extended in this article by defining a more complex pattern for computing similarity. The pattern is defined as the concatenated vector last $\Delta window$ hours of observations and single-value predictions, along with the next $\Delta t$ days of single-value predictions.   In this way, the pattern captures (a) recent agreement between the model and observation, indicating whether the model is well adjusted to the current physical system, as well as (b) the upcoming direction the model predicts. The upcoming direction the model predicts is critical because it matches against patterns, such as new high-speed streams, which carry higher uncertainty than others. 

In Figure \ref{fig:illustration}, we give an illustration of our probabilistic forecast around a stream interaction region reaching Earth in May 2010.  Stream interaction regions, where a high-speed stream trails a low-speed stream, are notably difficult for current solar wind models to accurately predict. \texthl{Though difficult, predicting them is crucial because they form regions of higher plasma density and stronger magnetic field, and can be a source of recurring space weather}. The top panel shows a reference observation from the Advanced Composition Explorer (ACE, \citet{chiu1998ace}) in red, the single-value ADAPT-WSA prediction in blue, and the probabilistic prediction as shaded regions. The darker shaded blue region indicates the 75\% percentile, and the lighter shaded blue region indicates the 95\% percentile. 

In this illustration, a key property of the probabilistic forecast is that the uncertainty adapts to the incoming stream interaction region, growing with it, and then returns to a lower level after the single-value ADAPT-WSA predictions and the observations begin agreeing again. The middle panel shows the width of the 75\% percentile, with arrows pointing to times that highlight this change. 

In the bottom panel, there are two probability distributions for the labeled \texthl{\textit{Time A} (2010-05-21, 12 UT) and \textit{Time B} (2010-05-27, 12 UT)} of the event. In \textit{Time A}, we see that the probabilistic prediction predicts a distribution mostly below the single-value prediction, which we attribute to the methodology's ability to apply an empirical bias correction obtained from the historical record using the fitted distribution. This can be understood to be because the single-value predictions indicate an incoming high-speed stream, but historical performance indicates that a high-speed stream of this magnitude is often a slight over-prediction. In \textit{Time B}, we see that no bias correction was necessary, but the distribution is skewed \texthl{with its tail extending towards the right}.

The organization of this article is as follows. In Section \ref{sec:adaptwsa}, we review the ADAPT-WSA and WSA point-parcel simulation methodology. In Section \ref{sec:dataset}, we describe the historical record of runs that drive the data-driven methods. In Section \ref{sec:formethod}, we describe the technical details of the methodology for making probabilistic predictions. Section \ref{sec:results} discusses calibration of free parameters, validation, and metrics for the technique, and some features of the observed uncertainty, such as the dominant time scales and solar cycle dependence. Finally, the conclusion is in Section \ref{sec:conclusion}.

\section{ADAPT-WSA Solar Wind Speed Model} \label{sec:adaptwsa}

\begin{figure}
    \centering
    \includegraphics[width=1\linewidth]{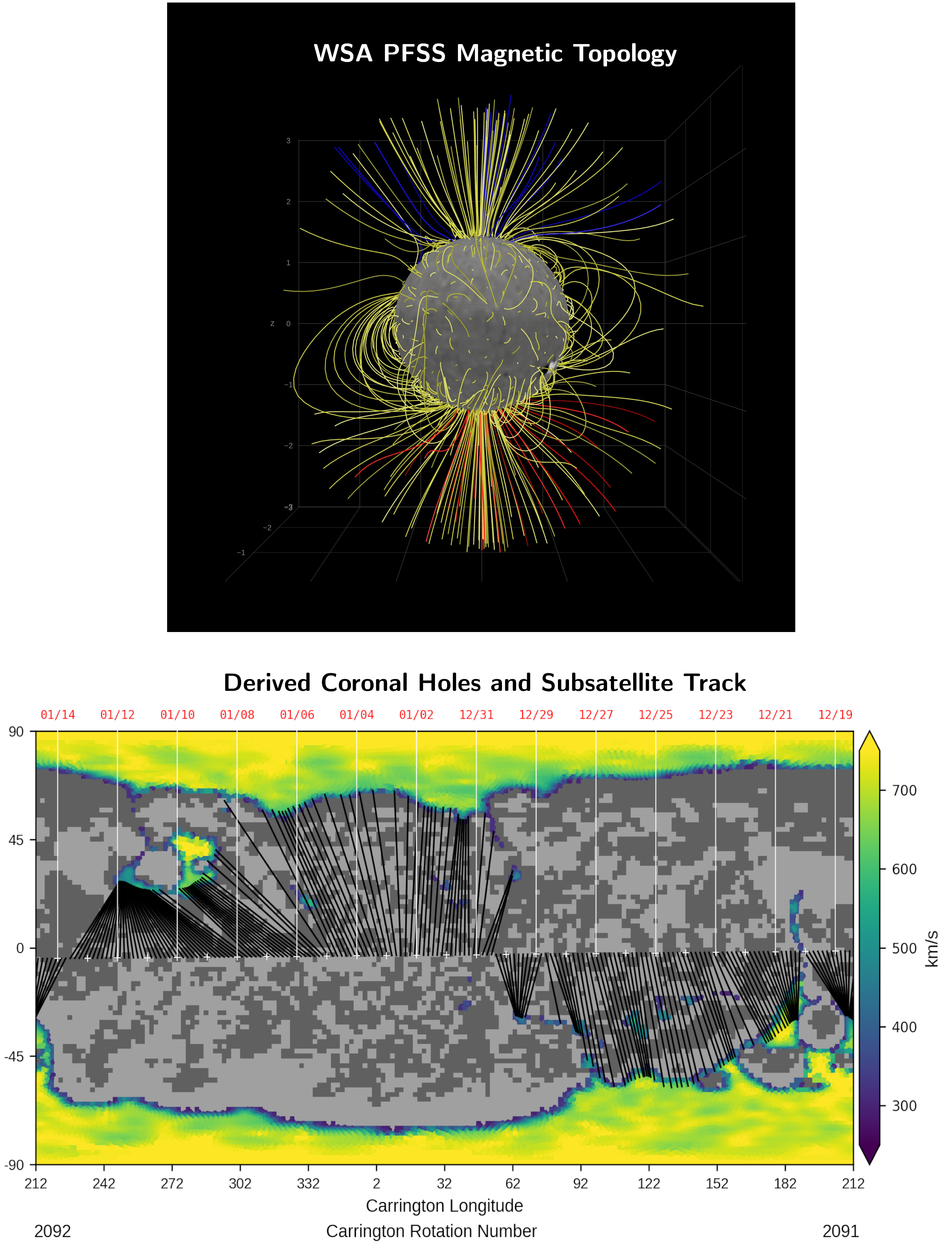}
    \caption{Modeling outputs from the WSA Solar Wind Model, displaying the magnetic topology \texthl{(top panel)} and the photospheric solar wind sources along the subsatellite track along with the speeds assigned within derived coronal holes \texthl{(bottom panel)}. \texthl{The bottom plot shows large coronal holes at poles, with some smaller coronal holes at lower latitudes. The black lines in the bottom panel connect the Earth subsatellite point with the coronal hole the parcel originated from.}}
    \label{fig:wsamodel}
\end{figure}

We use the ADAPT-WSA model with the WSA point-parcel simulation to model the solar wind flow into the heliosphere. Multiple options for the heliosphere models exist, from reduced physics models such as the built-in WSA point-parcel simulation \citep{da2023ensemble,wallace2022new,wallace2025connecting}, and HUXt \citep{barnard2022huxt,owens2020computationally}, to full three-dimensional magnetohydrodynamic (MHD) numerical simulations such as Enlil \citep{pizzo2011wang} and \texthl{the European Heliospheric Forecasting Information Asset (EUHFORIA;} \citealp{pomoell2018euhforia,hinterreiter2019assessing}). The main advantages of  reduced physics models are their computational speed and modest hardware requirements.

We address the modeling task using the WSA point-parcel simulation due to its simplicity and speed. We note that the methodology developed in this work can be readily applied, with minimal modification, to any other heliospheric model, provided a comparable simulation is prepared to provide a data-driven assessment of the overall forecasting pipeline performance (see the next section, \textit{Simulation Dataset and Observations}).

We utilize global photospheric magnetic field maps from ADAPT as input to WSA. ADAPT models the photospheric magnetic field state (as global maps) using a set of magnetic flux transport equations and ongoing data assimilation of near-side magnetogram measurements \citep{arge2010air}. Assimilated measurements may originate from instruments such as the \texthl{Global Oscillation Network Group (GONG;}  \citealp{plowman2020calibrating,plowman2020calibrating2}), the Solar Dynamics Observatory’s Helioseismic and Magnetic Imager  (SDO/HMI, \citet{hoeksema2014helioseismic}), or the Solar Orbiter Polarimetric and Helioseismic Imager (SolO/PHI, \citet{solanki2014polarimetric}). The ADAPT model generates an ensemble of solutions, representing a range of global magnetic field maps, each known as a single realization. Each ADAPT realization corresponds to different samplings of convection cell parameters \citep{hickmann2015data}. In this work, we use ADAPT driven by GONG measurements, as these maps are released with a higher time cadence than the other  variants.

To determine the three-dimensional magnetic field between $1~R_\odot$ and $5~R_\odot$, two potential field (PF) models are used. The inner model is a traditional potential field source surface (PFSS) model \citep{altschuler1969magnetic,wang1992potential}, with a spherical inner boundary of $1~R_\odot$, the photosphere, and $2.51~R_\odot$, the nominal source surface. By virtue of the model, every field line at the source surface is open and radial, making this a critical surface within the model. Between $2.49~R_\odot$ and $5~R_\odot$, the magnetic field is determined by the Schatten Current Sheet (SCS) \citep{schatten1971large} model. A small overlap region between the PFSS and SCS (2.49-2.51 $R_\odot$) is used to couple the PFSS and SCS domains. 

For the remainder of this section, we will describe the base ADAPT-WSA methodologies for predicting the solar wind speed at the spherical $5~R_\odot$ boundary, and the heliospheric WSA point-parcel simulation for propagating solar wind from that $5~R_\odot$ boundary condition to Earth. The methodology described here is the same as that used in WSA v5.2, as deployed at the Community Coordinate Modeling Center (CCMC).

The core methodology of the coupled WSA model to produce solar wind speeds at $5~R_\odot$ can be broken into two parts. \texthl{After determining the three-dimensional magnetic field between $1~R_\odot$ and $5~R_\odot$, WSA} analyzes the three-dimensional magnetic topology to produce values of \texthl{(a)} the coronal hole boundary distance, $\theta_b$, the minimum angular distance from the photospheric magnetic field footpoint of the traced field line to the nearest coronal boundary, and \texthl{(b)} the magnetic expansion factor, $f_s$, evaluated using the magnetic field values of the field line at $2.5~R_\odot$ (the source surface) and the photosphere. The values are produced for the footpoint of each field line starting at $5~R_\odot$. These two key variables are then utilized in the WSA empirical relationship,

\begin{equation} \label{eqn:wsaempirical}
    V(f_s, \theta_b) = 285 + \frac{625}{(1 + f_s)^{2/9}} \left\{ 1.0 - 0.8 e^{-(\frac{\theta_b}{2})^2} \right\}^{3}~\mathrm{km/s},
\end{equation}
where $V(f_s, \theta_b)$ is the predicted solar wind speed  \citep{riley2015role,wallace2020relationship,jones2024quantitative}.  The basic structure of the empirical relationship is to (a) assign faster speeds to solar wind flowing from deeper within the center of a coronal hole and (b) assign slower speeds near the boundary of coronal holes, where $f_s$ tends to be large.

 An illustration of the magnetic field lines in the PFSS region is shown in the \texthl{top} panel of Figure \ref{fig:wsamodel}. The \texthl{bottom} panel displays a map of the surface magnetic field polarity (background, gray) and the derived coronal holes (colored regions), colored by their out-flowing solar  wind speed assigned using the WSA empirical relationship (Equation \ref{eqn:wsaempirical}). The lines extending from around the equator correspond to the Earth subsatellite track, with the line pointing to the coronal hole from which the solar wind at Earth was modeled to originate.  

The WSA point-parcel simulation is a 1-D kinematic model that works by moving parcels quasi-ballistically outward from $5~R_\odot$ to their target location. \texthl{In this work, the target location is Earth, though more generally the model can be applied to other locations such as Mars and arbitrary satellite trajectories in the inner heliosphere.}  The simulation is parameterized by a time \texthl{delay} for the forecasting horizon, which generally varies between 1 and 7 days, denoted here as $\Delta t$, and a variable which defines how close a parcel must be to the target location, denoted here as $\delta t$. To make predictions for times $t+\Delta t$, parcels are started along the subsatellite track at a radius of $5~R_\odot$ using speeds assigned by the empirical relationship and the coronal solution created using an ADAPT map for time $t$.  Parcels are ballistically moved radially outward with no momentum exchange and subject to the restriction that fast parcels may not overtake slow parcels as a minimal effort to address stream interaction regions. \texthl{Because of this restriction, we choose to label the algorithm as quasi-ballistic instead of classically ballistic.}

The WSA point-parcel simulation does not capture coronal mass ejection (CME) physics because the underlying assumptions of the WSA model itself do not apply to active regions and CME reconnection/ejection. This limitation will be absorbed into the uncertainty predicted by the probabilistic \texthl{method} we develop.  \texthl{For models that do involve CME propagation, one could apply our method directly to a historical record of that model's performance, as our method would automatically adapt to CME  patterns in the data.}

Parcels are considered to reach the target location when they cross the section of Earth's ephemeris between $(t+\Delta t) - \delta t$ and $(t+\Delta t) + \delta t$. For this work, we use a $\delta t$ of 6 hours, and repeat the calculation for $\Delta t = 1, 2, ..., 7 ~\textrm{days}$.  \texthl{Selecting a large value of $\delta t$ results in more irregularly spaced points, but this is not always advantageous as the additional points arrive less precisely at the target location.}

\texthl{Compared to other solar wind propagation models such as HUXt, the WSA point-parcel simulation has the advantage that it makes it simple to carry forward information associated with the photospheric and coronal source regions. Such information includes the footpoint longitude, latitude, radial polarity, expansion factor, and coronal hole boundary distance. With this information carried forward, modeling tasks that would normally require tracing the flow field (such as determining the photospheric footpoint) become trivial.}

The WSA point-parcel simulation yields predictions of the solar wind at irregularly spaced times; in the next section, we will discuss binning procedures to create regularly spaced time-series datasets. While ADAPT-WSA provides multiple realizations per the 12-member ADAPT ensemble, we predict separate uncertainty values for each realization. For more information about the WSA point-parcel simulation can be found in \citet{da2023ensemble}.

\section{Simulation Dataset and Observations} \label{sec:dataset}
To create our data-driven approach to probabilistic forecasting, we produced a \texthl{11}-year-long run of ADAPT-WSA and the WSA point-parcel simulation. The simulation dataset spans between 2010 and 2020 inclusively, covering approximately one solar cycle. It is driven by the GONG network,  using four maps per day, spaced approximately 6 hours apart, and repeated for each of the ADAPT realizations and each days ahead values between 1 and 7 days.  This produces about 16,000 predictions each. These predictions are binned into 6-hour cadence to allow for pairing with observations. \texthl{While features of the solar wind exist on time scales under 6-hours, the resolution of the WSA model is only 2$^\circ$, which corresponds to a prediction capability for time scales around 3.6 hours when considering the solar rotation as the limiting factor ($27~\textrm{days} \times 2^\circ / 360^\circ = 3.6 \textrm{~hours}$). Because the WSA point parcel simulation's predictions are irregular, we chose a larger bin size to minimize the number of bins that do not contain a prediction. }

The solar wind speed observations are taken from Earth using the ACE spacecraft's Solar Wind Electron Proton Alpha Monitor (SWEPAM, \citealp{mccomas1998solar}) instrument. ACE observations used are \texthl{Level 2 data,} initially at hourly cadence,  averaged into the same 6-hour cadence as the predictions. 

The simulation dataset, including the complete WSA coronal solutions in FITS format, is made openly available with this manuscript. For information about data access, see the \textit{Open Research} section.

\section{Forecasting Methodology} \label{sec:formethod}
Conversations with forecasters involved in operations at the NOAA Space Weather Prediction Center revealed that a prevalent technique for assessing the current performance of a solar wind model involves two key  pieces of information. First, the recent agreement between the model's predictions and recent observations would indicate whether the model is likely to be at a strong point, where its assumptions hold, and whether the methodology is well-suited to the present time.  While things may indeed change, the trend is that if agreement is currently strong, it is likely to continue in the immediate near future, and this inspires a degree of trust.

The second piece of information is \textit{where the model is going}, i.e., the forecast in the future for which no observations are yet available. This information is shaped by the trajectory the model predicts; for solar wind speed, this may be a new high speed stream following a currently slow stream (a stream interaction region), a gradual decline in the current speed, or a steady continuation of the current speed. Each of these processes has its own associated uncertainty \citep{owens2008metrics,norquist2013forecast,jian2015validation,reiss2016verification}. The approach we take here is to use the historical record to capture the uncertainty for each of these patterns. 

We develop a probabilistic forecasting algorithm based on these two pieces of information. In a sense we use both principles: incorporation of recent observations, and data-driven modeling, by way of \texthl{leveraging} a historical record to describe model performance. 

Our method is a novel variant of analog ensembles. Analog ensembles in the sense of statistical theory are categorized as $k$-NN ($k$ nearest neighbors) algorithms.  In $k$-NN algorithms, a representative vector, denoted here as $\mathbf{u_t}$, is defined for the current time and for each time in a historical database. To perform inference at the current time, the set of $k$ closest neighbors, \texthl{as determined by the Euclidean distance}, is retrieved from the historical database and used to make the prediction. 


The action performed with the set of neighbors depends on the application of the $k$-NN algorithm; what is done here is a novel variation on the traditional methodology. Traditionally, $k$-NN algorithms are used for regression or classification. In regression, the predicted value may be a weighted average of neighbors; in classification, it may be the most common class. Notable examples of $k$-NN algorithms in Heliophysics include the AI-based Reconstruction of the Geospace Unified System (\texthl{ARGUS;}  \citealp{stephens2025empirical,stephens2019global,sitnov2008dynamical}) for magnetic field modeling within Earth's magnetosphere. 

We define $\mathbf{u}_t$ based on the three pieces of key information. Our definition uses recent observations, recent single-value predictions and upcoming single-value predictions from a single realization of ADAPT-WSA to construct a forecasting \textit{scenario}, and we use the neighbor lookup to data-mine comparable scenarios. To model uncertainty, we use the prediction errors in these comparable scenarios to empirically model a probability distribution for the solar wind speed at the current time. At the root, we are using neighbors found from matching similar scenarios $\mathbf{u}_t$, in order to  empirically build the probabilistic forecast denoted by $P(V~|~\mathbf{u}_t)$, where $V$ is the solar wind speed. 

\begin{figure}
    \centering
    \includegraphics[width=1\linewidth]{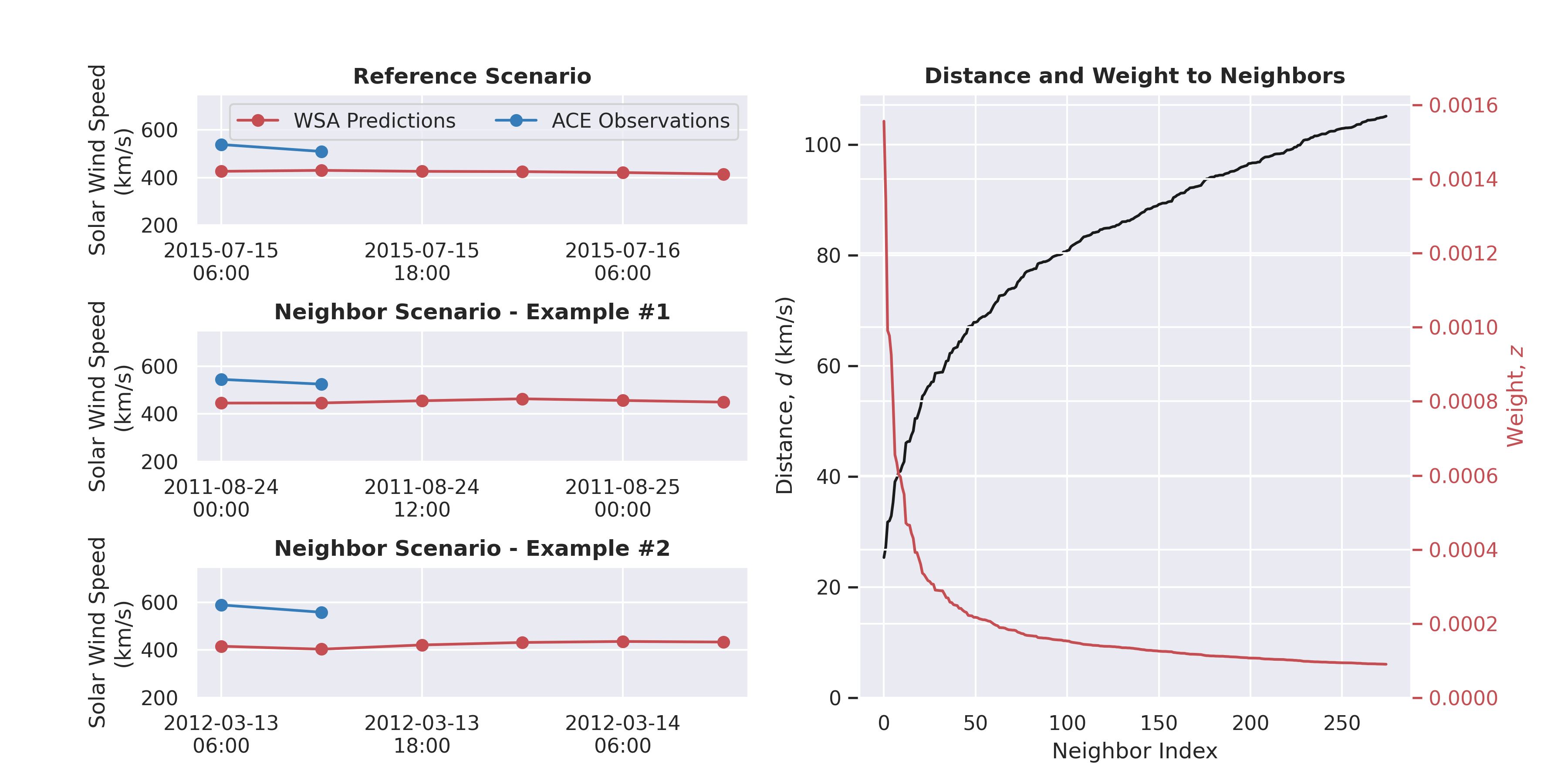}
    \caption{Illustration of a reference representative “scenario” vector (left panel \#1), and success finding similar scenarios (left panels \#2 and \#3) from other periods within the dataset of historical records (2010-2020). \texthl{The plot on the right shows the increase in distance to the neighbor (black line) and drop-off in weight (red line) as neighbor index increases. The first neighbors are closest distance-wise to the reference scenario, and are weighted more highly accordingly (unnormalized weight $z=1/d^2$). }}
        \label{fig:neighbors}

\end{figure}

The representative vector $\mathbf{u}_t$ for a given single-value forecast, parameterized by the number of days ahead $\Delta t$, is defined as the last 12 hours of observations, concatenated with the previous 12 hours of single-value predictions, concatenated with the following $\Delta t$ days of single-value predictions. Recent observations and predictions yield information about recent performance, and future predictions provide insight into where the model is headed. For information on how we selected the 12-hour window size, see Section \ref{sec:results}. In Figure \ref{fig:neighbors}, a plot of the recent observations, recent predictions, and future predictions for a sample time can be seen in the top left panel. In the two panels directly below, there are two neighbors (one in \texthl{August} 2011 and one in March 2012). These are instances from a different time when the forecasting scenario was very similar, and, by nature of being pulled from a historical database,  outcomes for the prediction effectiveness are available.   

We select $k=275$ neighbors out of about 16,000 possible neighbors for use in probability distribution modeling. We note that although many neighbors are selected, many of the selections are close in time, as the representative vectors of $u_i$ and $u_{i+1}$ are often very close. In the right panel of Figure \ref{fig:neighbors}, we show that the drop-off \texthl{in distance to} increasing neighbor indices, as defined by the distance $d = d(\mathbf{u}_t, \mathbf{u}_t')$, where \texthl{ $\mathbf{u}_t$ is state vector of the reference scenario and $\mathbf{u}_t'$ is the state vector of the neighbor.} The right panel shows the weight we assign to each neighbor based on this distance, which is \texthl{$z(\mathbf{u}_t, \mathbf{u}_t') = 1 / d(\mathbf{u}_t, \mathbf{u}_t')^2$}.

To model the probability distribution for the solar wind speed, we utilize a tool known as the skew normal distribution \citep{azzalini2005skew,yuan2018utilizing,allard2012skew,kim2020prefeasibility}. A skew normal distribution is a variation of the normal distribution with a location parameter, $\xi$, which acts like the normal mean parameter $\mu$, a scale parameter, $\omega$, which acts like the normal standard deviation parameter $\sigma$, and a third shape parameter, $\alpha$, which controls the degree of skewness.  We note to avoid confusion that $\xi$ is not in fact the mean; the mean is given by $\xi + \omega \frac{\alpha}{\sqrt{1 + \alpha^2}} \sqrt{\frac{2}{\pi}}$. Furthermore, the scale $\omega$ is not quite the standard deviation, and the shape parameter $\alpha$ is not quite the proper statistical measure of skewness.

\begin{figure}
    \centering
    \includegraphics[width=1\linewidth]{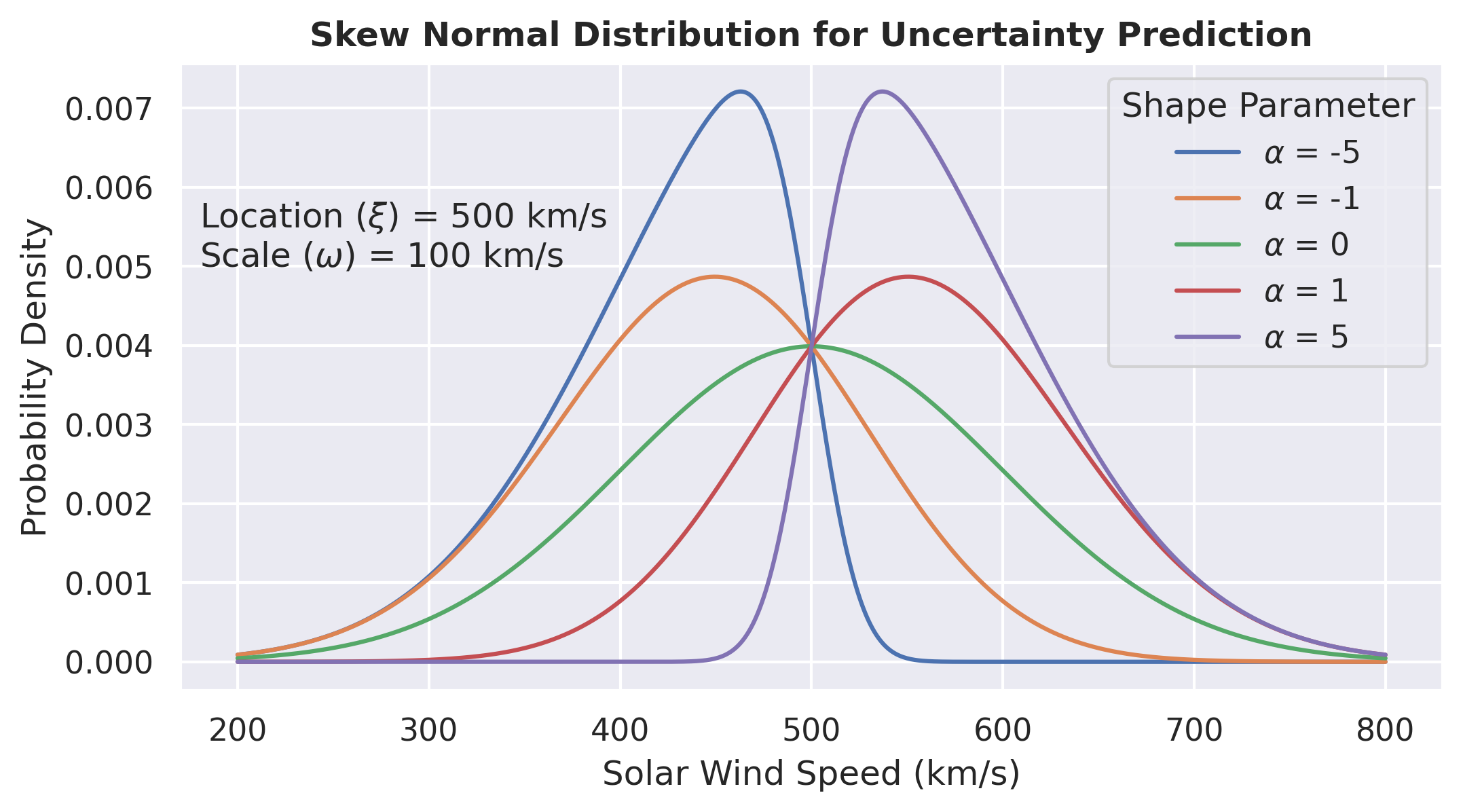}
    \caption{A graphical representation of skewed normal distributions for varying values of $\alpha$ with \texthl{fixed} location $\xi= 500~\mathrm{km/s}$ and scale $\omega=100~\mathrm{km/s}$. \texthl{The first two parameters of the skew distribution, $\xi$ and $\omega$, are conceptually analogous to the normal distribution mean ($\mu$)  and standard deviation ($\sigma$) in terms of controlling its shape. The third parameters $\alpha$ is a new parameter which controls skew and asymmetry.  We note that though $\xi$ and $\omega$ have similar effects for shaping the distribution, $\xi$ is not precisely the mean, $\sigma$ not precisely the standard deviation, and $\alpha$ does not follow  the formal definition of statistical skew.}}
    \label{fig:skewnorm}
\end{figure}

The probability density function for the skew normal distribution of the solar wind speed variable $V$ is given by,
\begin{equation}
    f(V) = 2 \phi(V) \Phi(\alpha V),
\end{equation}
where,
\begin{equation}
    \phi(V) = \frac{1}{\omega\sqrt{2\pi}} e^{-\frac{1}{2}(\frac{V - \xi}{\omega})^2},
\end{equation}
is the standard normal probability density function, and $\Phi(V)$ is the normal cumulative distribution function. In this way, values of $\alpha > 0$ ($\alpha < 0$) generate a skewed tail towards higher (lower) values, and a value of $\alpha = 0$ yields the normal distribution identically. A graphical representation of skewed normal distributions for varying values of $\alpha$ with location $\xi= 500~\mathrm{km/s}$ and scale $\omega=100~\mathrm{km/s}$ is shown in Figure \ref{fig:skewnorm}.

Our reason for using a skew normal distribution over a normal distribution is that it allows asymmetric uncertainty. This is important because solar wind speed largely exists within a bounded interval. For instance, if the predicted solar wind speed is 200 km/s, it is very unlikely the true speed is appreciably  lower,  and therefore having an error bar below that is as large below as it is above is statistically statistically unrealistic. We would expect that asymmetric error bar modeling be appropriate for other models as well. Furthermore, \texthl{when using a skew normal distribution}, our validation metrics (to be discussed later) improved greatly. To predict the solar wind speed distribution at a given time $t+\Delta t$, we collect the errors from each neighbor for each neighbor $i$,
\begin{equation}
    \varepsilon_i=V_i^{obs}(t_i+\Delta t)-V_{i}^{pred}(t_i+\Delta t),
    \end{equation}
and add them relative to the nominal predicted solar wind speed at the target time $V_{pred}$ from the current time,
\begin{equation}
        V^{fitdata}_i = V^{pred}(t + \Delta t) + \varepsilon_i.
\end{equation}

These values are fitted with weighted maximum likelihood estimation (MLE) and the Limited-memory Broyden–Fletcher–Goldfarb–Shanno algorithm with bounds (L-BFGS-B, \citealp{byrd1995limited,zhu1997algorithm,virtanen2020scipy}). \texthl{This algorithm was chosen for its ease of use and effectiveness at solving optimization problems without explicitly defined gradients.} This approach frames the fit as a weighted optimization problem to produce the skew normal parameters $\xi$, $\omega$, and $\alpha$, which are most likely to yield the given fit data as samples. Without weights, it aims to maximize the joint probability $\prod f(V_i^{fitdata}; \xi, \omega, \alpha)$ by minimizing the negative of the more numerically stable logarithm (recall $\mathrm{log}(\prod X_i)=\sum \mathrm{log}(X_i)$). With weights introduced (\texthl{$\tilde{z}_i$, a normalized version of $z_i$ such that the sum of the weights equals 1}), the function  $L(\xi, \omega, \alpha)$ being minimized is,
\begin{equation}
    L(\xi, \omega, \alpha) =  -\sum \tilde{z}_i \mathrm{log}(f(V_i^{fitdata}; \xi, \omega, \alpha)).
\end{equation}
The optimization was done using bounds of  [$10^{-6}$, $\infty$) for $\omega$, and [-20, 20] for $\alpha$. Initial values were set using the results of an unweighted MLE fit to the skew normal distribution. Pseudo-code for the whole algorithm can be found in \ref{apdx:pseudocode}.

By fitting a skewed distribution with three parameters, we can adjust for empirically observed biases (through $\xi$), the average scale of errors (through $\omega$), and whether the error bar should be larger above or below the median value (through $\alpha$). 

Revisiting Figure \ref{fig:illustration} (top panel), the distribution shown for \textit{Time A} illustrates a distribution that states an overestimating bias by placing the center of the mostly skewed distribution slightly below the prediction.  This says that, on average, a predicted new high-speed stream of this type in similar scenarios overestimates the observations by around 100 km/s, and the most appropriate adjustment is a bias correction by that amount, with the observations falling roughly in equal parts above and below the bias-corrected value.   

The distribution shown for \textit{Time B} illustrates a distribution that does not need a bias adjustment. Still, it benefits from the introduction of a skew, with the error bar larger above the predicted value than below. The predicted solar wind speed at this time is already slow, and it makes sense that if the prediction were wrong, it would most likely be faster. 

\texthl{We remind the reader that we did not remove CMEs from the observation time series; this was done intentionally. By not removing CMEs, we obtain error bars that, by design, adapt in shape to the presence of CMEs, extending the distribution to include the possibility of a CME. }

\section{Results and Discussion} \label{sec:results}
To select the values of the number of neighbors, $k=275$, and the recency window of observations, $\Delta window=\textrm{12~hrs}$, we perform an exhaustive search of the parameter space using a custom metric.

It is desirable for a probabilistic prediction to have the property that for a given $p\%$ percentile and the corresponding  bounds determined by the algorithm, observations lie within these bounds approximately $p\%$ of the time. For example, the observations should fall within the the 50th percentile about half of the time.  We call this concept the percentile efficiency.

\begin{table}
    \centering
    \begin{tabular}{cccccccc}\toprule
         \texthl{Target} Percentile&  \multicolumn{7}{c}{Rate Observations Fell within Predicted Percentile (Days Ahead)}\\\midrule
         &  
1 Day&  2 Days&  3 Days&  4 Days&  5 Days&  6 Days& 7 Days\\
\bottomrule
         25\%&  26.3\%&  26.1\%&  25.8\%&  25.8\%&  25.7\%&  25.7\%& 25.6\%
\\
         50\%&  50.6\%&  50.2\%&  50.3\%&  50.5\%&  50.4\%&  50.0\%& 50.0\%\\
         75\%&  75.1\%&  74.1\%&  74.3\%&  74.9\%&  74.3\%&  75.0\%& 72.8\%
\\ \bottomrule
    \end{tabular}
    \caption{Rate at which observations from ACE fell within the predicted percentiles using the method of the manuscript. \texthl{In a perfect model, the reported rates would be identical to the target percentile in the left column. Here, they are close, generally under 1\% for most cases.} Statistics are taken from ADAPT Realization 0 at Earth.}
    \label{tab:percentiles}
\end{table}

Percentile efficiencies for the selected $k=275$ and $\Delta window=\textrm{12~hrs}$ from ADAPT realization 0 predictions at Earth are shown in Table \ref{tab:percentiles}. When we calculate the percentile efficiency, we run the algorithm across the dataset from Section \ref{sec:dataset} \texthl{with} the notable distinction that neighbors within $\pm1$ Carrington rotation be excluded from use as neighbors. This is an ad hoc method for effectively creating a unique train/test split at each timestep, satisfying the requirement of not testing on the training data while maximizing the available data. 

To aggregate percentile efficiency across all percentiles, we define a new concept, the Total Percentile Score ($TPS$). For a target percentile $P_{target}$, between 0 and 100, we define the rate of observations that fall within the predicted bounds as $P_{obs}(P_{target})$. The metric which summarizes the \textit{percentile score} over the entire range is therefore given by,

\begin{equation} \label{eqn:tps}
    TPS = \int_{P_{target}=0}^{100}{|P_{target} - P_{obs}(P_{target})|dP_{target}},
\end{equation}
where $TPS$ is the total percentile score using predictions from a single day ahead value ($\Delta t$). We note that we consider the percentiles to be defined between 0\% and 100\%, though redefining them as fractions between 0 and 1 is straightforward. An intuitive way to understand the $TPS$ definition is that when the rate observations fall within the predicted bounds is off by $2.5\%$ at each percentile, then the $TPS= 100\times 2.5  = 250$.  

To search the values of $k$ and $\Delta window$, we calculated the sum of the $TPS$ values from each days ahead value between 1 and 7 days.  We then selected the lowest value. The parameter space searched was $k$ between 25 and 500 in increments of 25, and $\Delta window$ between 0 and 5 days in increments of 6 hours (the bin size). To make the problem computationally tractable, we searched using results from realization 0 and later confirmed, through data analysis, that the results were similar across the other realizations. \texthl{For more information and visualizations from the selection process,  see} \ref{apdx:optimalselection}.

It is helpful for statistical learning algorithms in the literature to compare their effectiveness against a simple baseline. This is useful, at a minimum, to demonstrate that the use of advanced analytical techniques over a more straightforward methodology is warranted. For this work, we compare our percentile scores against the simple baseline of treating the probabilistic prediction as a normal distribution centered on the WSA point-parcel simulation prediction, with a standard deviation set by the root mean square error. \texthl{The equation for $\sigma_{baseline}$ set this way is,}

\begin{equation}
    \sigma_{baseline} = \sqrt{\frac{1}{N} \sum_{i=1}^{N} {(V^{obs}(t_i)-V^{pred}(t_i))^2}},
\end{equation}
\texthl{where $N$ is the total number of paired points in the historical record}. In this simple baseline, $\sigma_{baseline}$ is static and does not vary as a function of time. The baseline is meant to capture a data-driven, albeit non-time-adaptive, method of attaching uncertainty to the WSA point-parcel simulation predictions without any advanced methodology.

\begin{figure}
    \centering
    \includegraphics[width=1\linewidth]{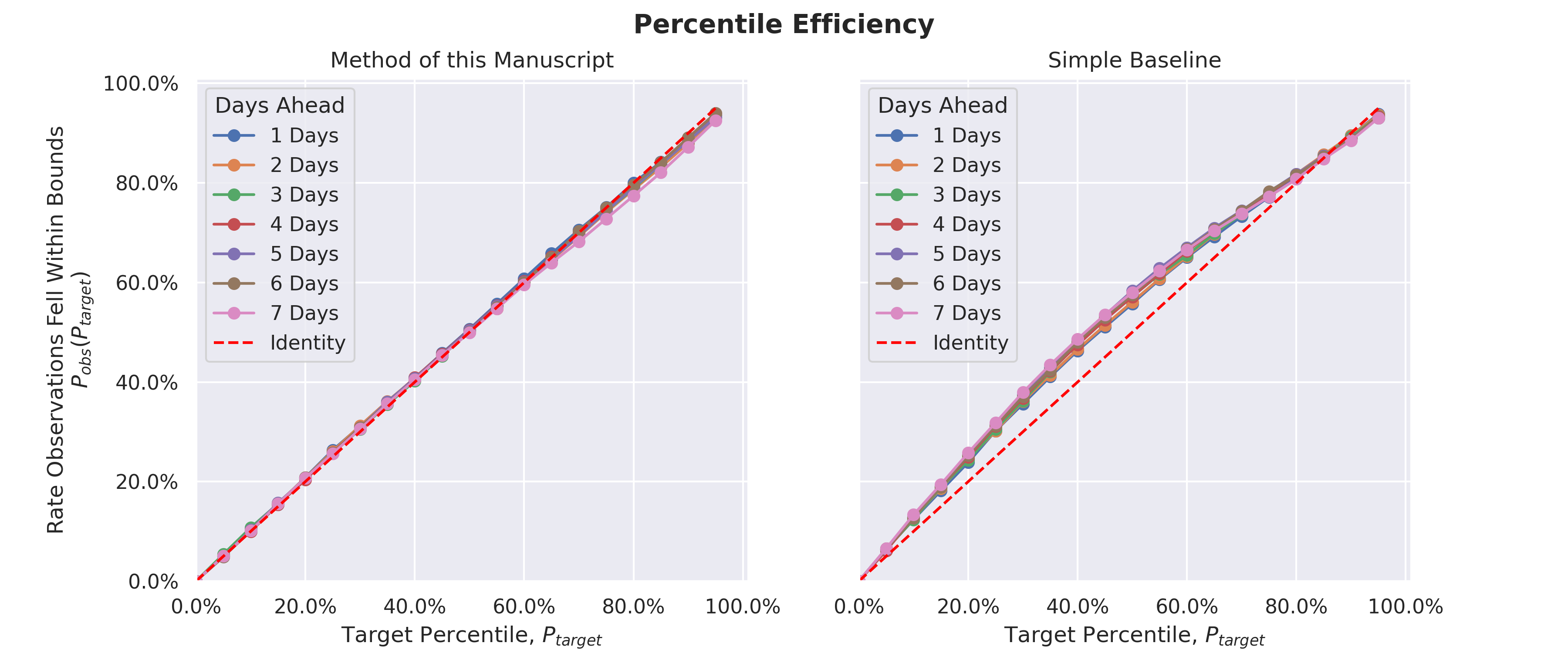}
    \caption{Comparison of Percentile Efficiency (what percentage of observations fall within the predicted percentile range) for the model and simple baseline. \texthl{The x-axis shows the target percentile for the error bars at each timestep. The y-axis axis shows the rate at which observations fell within those predicted error bars. For a flawless model, the line would be a perfect diagonal. Here, the method of this manuscript (left panel) has much better alignment to the perfect case than the simple baseline (right panel).}}
    \label{fig:percentile}
\end{figure}

In Figure \ref{fig:percentile}, we show comparisons of the target percentile, $P_{target}$, on the x-axes and the rate at which observations fell within the predicted bounds,  $P_{obs}(P_{target})$, on the y-axes, for both the method of this manuscript and the simple baseline. An ideal method would have both equal (leading to $TPS= 0$), yielding a straight line on the diagonal. This is nearly the case with our model. The right panel, which uses the simple baseline, shows much stronger deviations, especially around the $50\%$ mark, visually indicating that the method in this manuscript outperforms the simple baseline.

\begin{table}
    \centering
    \begin{tabular}{cccccccc}\toprule
 & \multicolumn{7}{c}{Total Percentile Scores (Days Ahead)}\\\midrule
         &  1 Day&  2 Days&  3 Days&  4 Days&  5 Days&  6 Days& 7 Days\\
         \bottomrule
         Our Model&  66.1&  82.4&  60.5&  50.0&  67.6&  42.3& 104.0\\
         Simple Baseline&  346.6&  378.5&  406.5&  426.3&  473.2&  449.8& 467.3
\\ \bottomrule
    \end{tabular}
    \caption{Total Percentile Scores (lower is better) for our model compared to the simple baseline described in the manuscript. \texthl{This table indicates that based on the total percentile score (defined in Equation} \ref{eqn:tps}),  \texthl{our model beats the simple baseline.}}.
    \label{tab:baseline}
\end{table}

A more quantitative assessment of improvement relative to the simple baseline is presented in Table \ref{tab:baseline}.  The table displays the $TPS$ for each day ahead (columns) for each of our model and the simple baseline (rows). The consistently lower  $TPS$ scores for our model indicate a quantitatively demonstrated improvement over the simple baseline. 

\begin{table}
    \centering
    \begin{tabular}{ccccccccc}\toprule
 & \multicolumn{8}{c}{\texthl{RMSE} (km/s, days ahead)}\\\midrule
         & Ongoing Solution &  1 Day&  2 Days&  3 Days&  4 Days&  5 Days&  6 Days& 7 Days\\
         \bottomrule
         WSA Point Parcel&N/A&  104.69&  103.33&  103.49&  104.99&  107.69&  108.74& 114.67\\
         Mean of Skew Normal&N/A&  71.47&  83.69&  87.26&  90.07&  92.57&  94.45& 99.61\\
 Median of Skew Normal &N/A& 72.18& 84.56& 88.17& 90.80& 93.68& 95.16&100.27\\
 \texthl{$\approx 1$ Rotation Recurrence} & \texthl{94.71}&N/A&N/A&N/A&N/A&N/A&N/A&N/A \\
 \texthl{WSA-Enlil$^\dagger$}& \texthl{97.2}&N/A&N/A&N/A&N/A&N/A&N/A&N/A \\
 \texthl{CORHEL$^\dagger$}  & \texthl{116}&N/A&N/A&N/A&N/A&N/A&N/A&N/A

    \end{tabular}
    \caption{Root-Mean-Square Error (\texthl{RMSE}) metric for treating the mean/median of the predicted skew normal distribution as a single-value time series forecast. WSA and Skew Normal parameters taken from realization 0. \texthl{$^\dagger$ Numbers reprinted from}  \citet{owens2008metrics} \texthl{ for comparison; it should be noted these are unmodified from their original 1-hour binning and were obtained from a different time period.}}
    \label{tab:rmse}
\end{table}

When the mean and median of the skew normal distribution, \texthl{calculated in a time-dependent way}, are used as single-value predictions, they outperform ADAPT-WSA point parcel predictions on the root-mean-square error (RMSE) metric. Used this way, the methodology essentially amounts to a post-processing \texthl{adjustment on the single-value prediction using recent observations. Applying this technique improves the RMSE metric. When adjusting using the \textit{mean} of the skew normal distribution for 3 day ahead predictions, the RMSE dropped from  103.49 km/s to 87.26 km/s. When adjusting using the \textit{median}, it dropped from 103.49 km/s to 88.17 km/s. We expect that the value added from this technique is the empirical correction of systematic over-estimation and under-estimation from the ADAPT-WSA model.}


The obtained RMSE scores are generally on the scale of those from \texthl{other physics-based models}  \citep{owens2008metrics}.  These RMSE scores exceed those reported in the literature for solar wind predictions based on machine learning  \citep{brown2022attention,raju2021cnn,upendran2020solar}, though this is an apples-to-oranges comparison regarding the approach proposed in the article, as  this method could be applied to each of those machine learning methods, possibly even yielding an improvement in RMSE as it does with ADAPT-WSA,  

A full report of RMSE scores for these single-value predictions is available in Table \ref{tab:rmse}. \texthl{Included in this table is a row labeled \textit{$\approx$1 Rotation Recurrence} for comparison, generated using ACE observations averaged into the same 6-hour bins and paired with observations $4\times27=108$ bins prior. Rows are included for two additional forecasting systems: WSA-Enlil  and CORona-HELiosphere (CORHEL;} \citealp{linker2009corhel}). \texthl{These systems produce ongoing solutions of the solar wind and are not parameterized by a fixed number of days ahead. The RMSE scores for WSA-Enlil and CORHEL are reprinted from} \citet{owens2008metrics} \texthl{for comparison; it is worth noting that they are unmodified from their original 1-hour binning and cover a different time period than studied here.}

\texthl{Comparing the RMSE scores to the \textit{$\approx$1 Rotation Recurrence} baseline, we notice that scores for the WSA point parcel simulation, WSA-Enlil, and CORHEL do not outperform the baseline. This is not a new finding; it is also reported in} \citep{owens2008metrics} \texthl{with cases also reported in} \citet{norquist2010comparative}. \texthl{The presence of this for WSA-Enlil has been shown to vary throughout a solar cycle} \citep{owens2005event}.

\texthl{While the base WSA point parcel, WSA-Enlil, and CORHEL do not beat recurrence, a notable improvement of this method is that combining the WSA point parcel simulation with the mean or median of the skew normal distribution does in fact beat persistence for up to 5-day ahead forecasts. Applying the method to WSA-Enlil, which on average performs better than the WSA point parcel simulation, may lead to even better results.}

\begin{figure}
    \centering
    \includegraphics[width=1\linewidth]{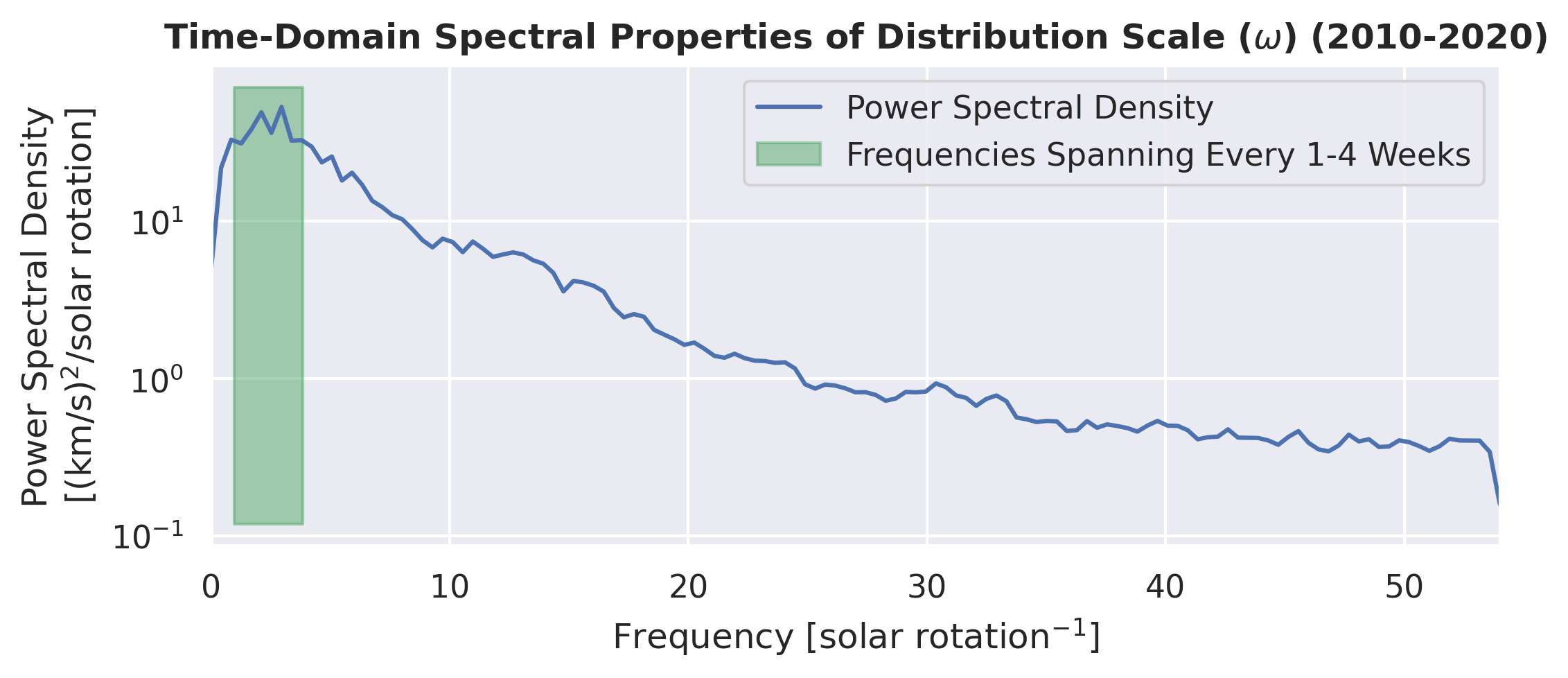}
    \caption{Time-domain spectral characteristics of the uncertainty, displayed through the Power Spectral Density (PSD). The peak frequency is twice per solar rotation (the time required for a feature to rotate across the visible disk of the sun), \texthl{with the broader peak corresponding to once every 1-4 weeks}. Data is taken from predictions at Earth using the ADAPT realization 0 and 3-day-ahead predictions. }
    \label{fig:spectral}
\end{figure}

We analyze patterns in the probabilistic forecasts \texthl{with the intent of understanding the model better}. After computing probabilistic estimates for the \texthl{11} year period in question using the same withholding methods as before, we perform a spectral analysis in the time domain of the forecast scale parameter ($\omega$) to test the dominant frequencies present (Figure \ref{fig:spectral}).  We notice dominant frequencies in the range of 1-4 weeks, which correspond to the time scales of a Carrington rotation (on the upper limit), and how long solar wind speeds persist from a single coronal hole source (lower limit). \texthl{We note that while one per solar rotation is within the range of the dominant frequencies, it is not an extreme spike. We expect a blend of frequencies around this range to be pronounced because of an aforementioned variety of features whose influence may come and go on shorter time scales}. This analysis implies the model is producing patterns in the uncertainty consistent with operational forecasting experience and knowledge of the underlying physical system.

\begin{figure}
    \centering
    \includegraphics[width=1\linewidth]{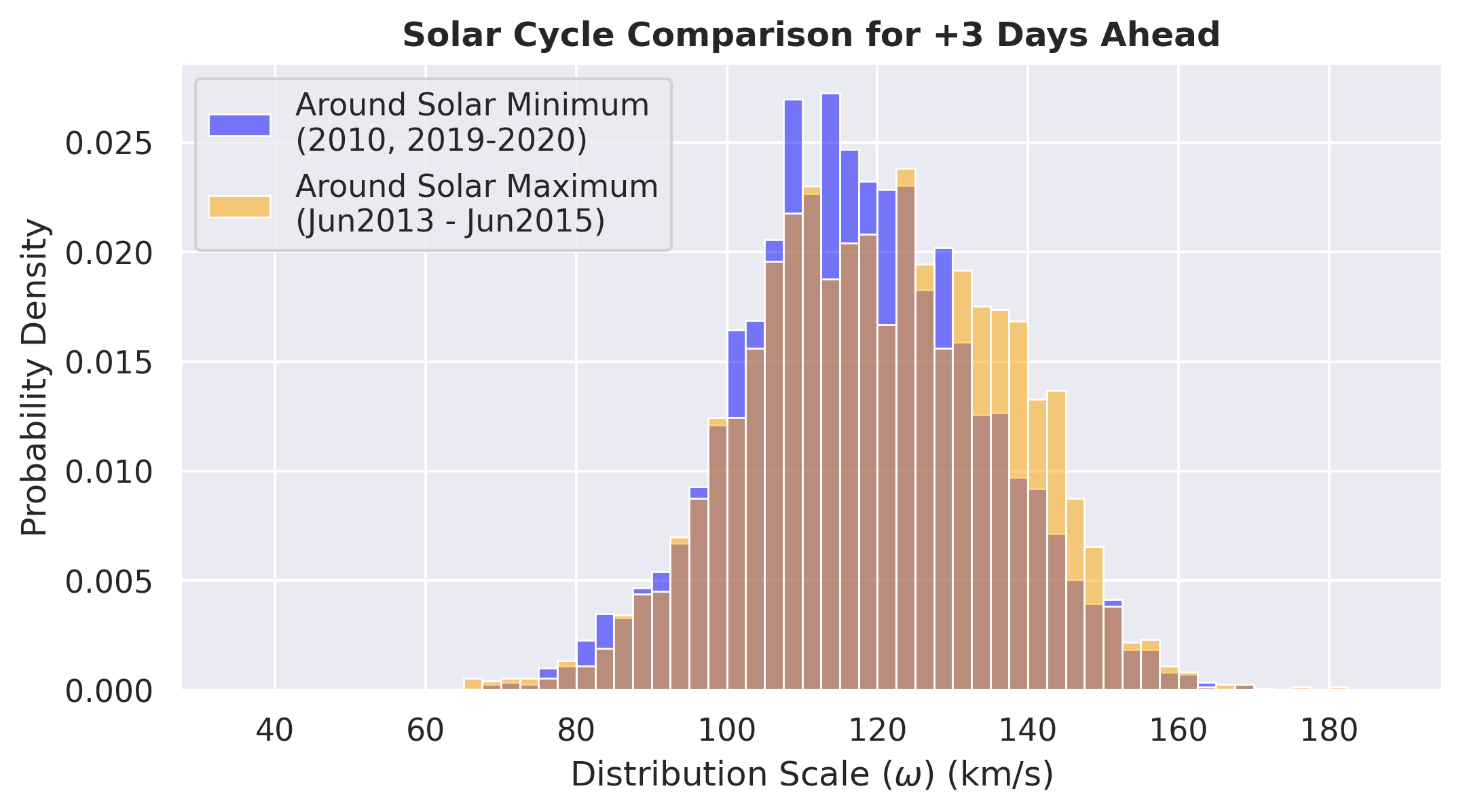}
    \caption{Histogram of distribution scales ($\omega$) for 3-day-ahead predictions with ADAPT realization 0 at Earth. The uncertainties predicted during solar maximum are overall slightly larger, \texthl{which is expected due to effects of increased coronal mass ejections  and limitations of PFSS models with more complex coronal field configurations. This plot reinforces that the model is able to capture solar cycle patterns through its k-NN methodology.}}
    \label{fig:solarcycle}
\end{figure}

In Figure \ref{fig:solarcycle}, we similarly analyze histograms of the scale parameter ($\omega$) for times around solar minimum (2010, 2019-2020) and solar maximum (June 2013-June 2015). The histogram y-axis is normalized to remove \texthl{any} effects from \texthl{a slight uneven number of points between the solar minimum and maximum datasets being considered}. We see slightly larger scales produced during solar maximum, consistent with the understanding of WSA and PFSS model performance that the solar wind speed is harder to predict during solar maximum in part due to the presence of coronal mass ejections \citep{reiss2016verification}.


\section{Conclusion and Summary} \label{sec:conclusion}
Probabilistic models and models with uncertainty quantification offer additional rich information for risk assessment and physical interpretation. This is especially true when the uncertainty can be linked to known limitations of the underlying model, such as broken physical assumptions and relatively complex phenomena. 

In this work, we developed a novel method extension of analog ensembles and $k$-NN for turning a single-value forecast of the solar wind speed into a probabilistic forecast. Our methodology leverages two key pieces of information, often used in practice by space weather forecasters: (a) the recent agreement between the single-value forecast and available observation, and (b) the upcoming set of future predictions and the related features there-in.  The methodology was validated  \texthl{to confirm that when the bounds of a $p^{th}$ percentile is predicted, observations fall within those bounds very close to $p\%$ of the time, with generally less than 1\% error} (see  Table \ref{tab:percentiles}).

The model uses the $k$-NN approach of referencing forecasting similar scenarios, as defined by those two pieces of information. It  is an interpretable method that leaves a direct link between the probabilistic forecast at a given time and the data from the historical record used to produce it. The methodology produces the probabilistic distribution as a three-parameter skew normal distribution. While non-parametric distributions could be created, using a three-parameter distribution has advantages operationally for its simplicity. Disadvantages of the methodology lie around the requirement of a large historical record of model performance, which may be difficult to obtain for compute-intensive forecasting pipelines such as those that rely on MHD simulation.  Analysis in Section \ref{sec:results} discovered features of the probabilistic forecasts in aggregate, namely solar cycle variation and the  dominant time scales of change in uncertainty scale. These features are consistent with knowledge of the solar wind forecasting problem as a whole, yet were derived without being explicitly cast into the model.

The method of this article adds calibrated probabilistic forecasts by building on top of an existing single-value forecasting pipeline as a post-processing step. The advantages over the single-value pipeline are the added information from predicting a distribution and the empirical bias correction, evident from the fact that using the mean of the predicted distribution yields smaller forecasting errors than using the single-value prediction itself.  Integrating recent solar wind observations into the ADAPT-WSA model directly is not straightforward, as the recent solar wind observations would have \texthl{influenced} the selected magnetic topology and the way to do this is not straight forward. By adding this information in a post-processing stage, the information from recent solar wind is integrated statistically without the need to modify the underlying physics of the forecasting pipeline. 

This gives rise to the potential for a family of post-processing algorithms, which improve forecasting results without modifying the internals of the baseline forecasting pipeline. This approach could provide a relatively simple way to reduce forecasting error even as the underlying models continue to improve.  By  means of statistically integrating new information and empirically calibrating away the model's own systematic biases, there is much room for improvement in this vein. 

\appendix
\section{Algorithm Pseudo-code} \label{apdx:pseudocode}
\begin{figure}
    \centering
    \includegraphics[width=1\linewidth]{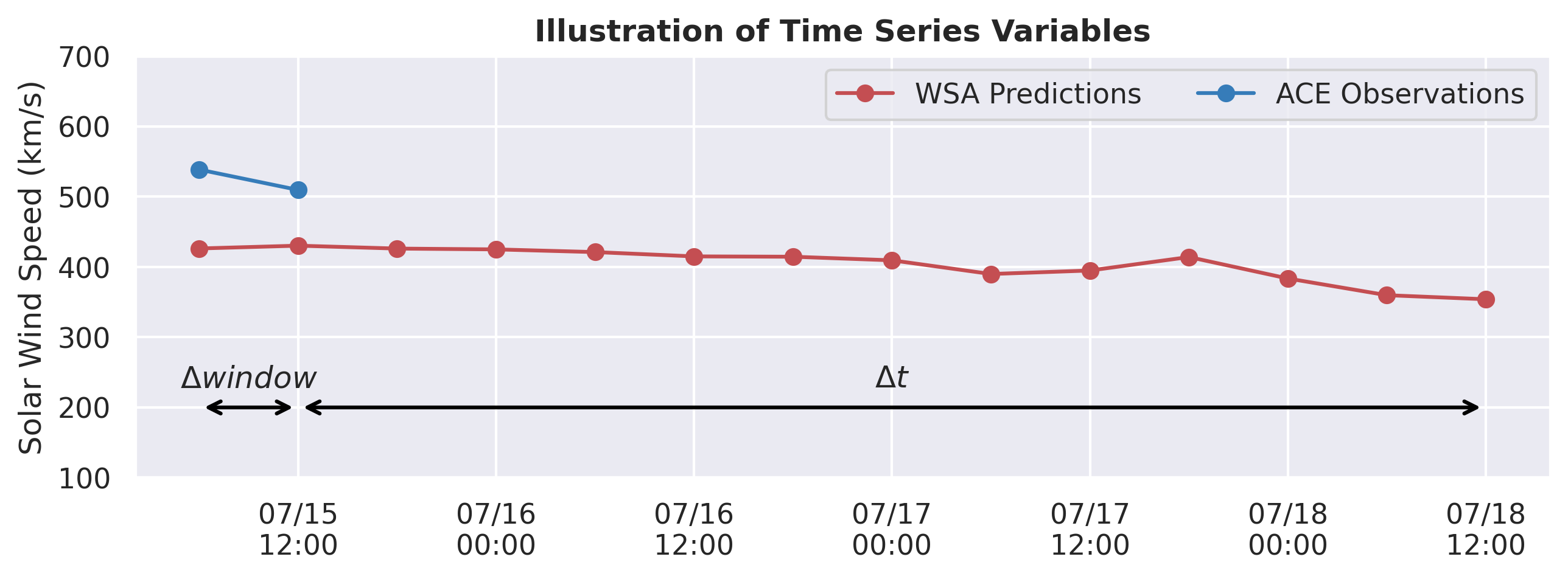}
    \caption{Illustration of variables $\Delta window$ (window size to track recent prediction/observation agreement) and $\Delta t$ (number of days ahead) used in the algorithm pseudocode. \texthl{In this plot, $\Delta t$ corresponds to 3 days ahead, but in other forecasting modes may take on values as low as 1 day or as high as 7 days. The number of points $\Delta window$ was selected based on a search over the parameter space, discussed in Section} \ref{sec:formethod}.}
    \label{fig:variables}
\end{figure}

Pseudo-code for the forecasting algorithm using $k$ nearest neighbors can be found below. The algorithm first looks up neighbors, stores errors relative to the current prediction, and performs a weighted fit to the skew normal distribution. Figure \ref{fig:variables} illustrates the parts of a representative vector, and labels the $\Delta window$ and $\Delta t$ variables (for $\Delta t=3~\textrm{days}$).

\begin{algorithm}
\caption{Generate a Probability Distribution with k-NN and Skew Normal Fit}\label{alg:cap}
\begin{algorithmic}
\\

\Comment{Lookup Neighbors}
\State $\mathrm{rep\_vec} \gets \mathrm{concatenate}(
    V^{obs}[t-\Delta \mathrm{window} : t],~
    V^{pred}[t-\Delta \mathrm{window} : t+\Delta t]
)$
\State $\mathrm{distances, ~neighbors} \gets \mathrm{knn\_lookup}(\mathrm{rep\_vec},~ k)$ 
\State $V^{fitdata}  \gets \mathrm{new\_array}(size=k)$
\\
\For{$i = 1$ \textbf{to} $k$} \Comment{Store Errors Relative to Current Prediction}
    \State $\varepsilon \gets  V^{obs}_i[t_i + \Delta t] - V^{pred}_i[t_i + \Delta t])$
    \State $V^{fitdata}[i] \gets  V^{pred}[t+\Delta t] + \varepsilon$
\EndFor
\\
\State $\mathrm{weights} \gets 1 /\mathrm{ distances}^2$ \Comment{Do Fit}
\State $\mathrm{location, ~scale, ~shape} \gets \mathrm{weighted\_skewnorm\_fit}(V^{fitdata},~ \mathrm{weights})$ 
\end{algorithmic}
\end{algorithm}

\section{Selection of Optimal Values for $k$ and $\Delta window$} \label{apdx:optimalselection} 

\begin{figure}
    \centering
    \includegraphics[width=1\linewidth]{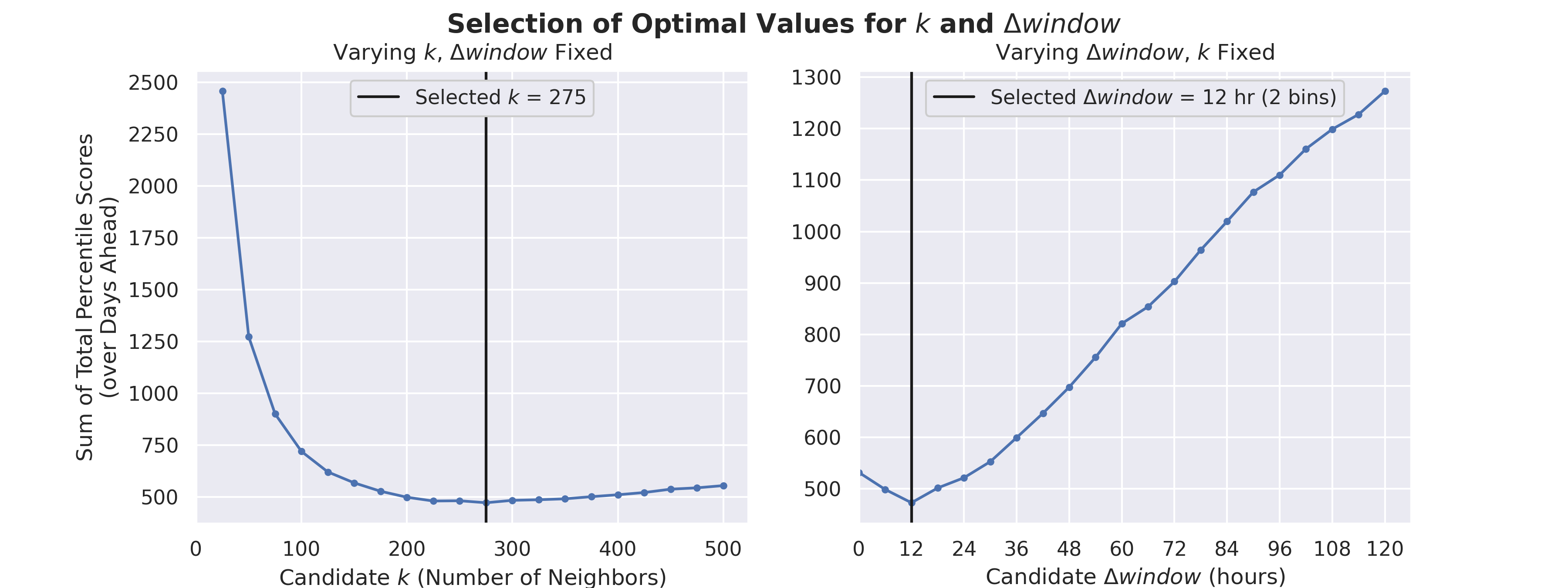}
    \caption{\texthl{This figure shows 1D cuts of the 2D optimization problem to select $k$ and $\Delta window$, around the selected values of $k=275$ and $\Delta window=12~\textrm{hours}$ (2 bins). The y-axis is the sum of the total percentile scores (Equation} \ref{eqn:tps}\texthl{) for days ahead values between 1 and 7 days. The full 2D grid searched was $k$ between 25 and 500 (inclusively, in increments of 25), and $\Delta window$ between 0 and 5 days (inclusively, in increments of the bin size, 6 hours).}}
    \label{fig:variables}
\end{figure}

\texthl{In this appendix section we present a plot showing a 2D cut of the exhaustive search process to find optimal values for $k$ and $\Delta window$. The final selected values were $k=275$ and $\Delta window=12~\textrm{hours}$ (2 bins), which appear on these plots as local minima. The function minimized is the sum of the total percentile score (TPS, Equation} \ref{eqn:tps}\texthl{) for days ahead values between 1 day and 7 days. }

\texthl{Intuitively, we understand the trade-off being evaluated are as follows. When $k$ is too small, there is insufficient information to model the distribution, but when $k$ is too large, fitting of the distribution is burdened by including neighbors which are less relevant. Similarly, when $\Delta window$ is too small, there is insufficient recent observational information being incorporated into the neighbor search, but when $\Delta window$ is too large, finding comparable neighbors is more difficult given the length of the historical record used in the study. It is expected that these values would change under different binning schemes and for longer duration historical records.}

\section*{Open Research Section}
Data used, including the 2010-2020 simulation dataset, ACE solar wind data, the processed k-NN dataset, and intermediary data analysis output is available on HuggingFace at \url{https://huggingface.co/datasets/ddasilva/probabilistic-solar-wind}. \textit{Data will also be uploaded to Zenodo after acceptance, but is not uploaded now because Zenodo will freeze the files.} The WSA coronal output in FITS format from the 2010-2020 simulation dataset is too large for Zenodo and is only available on HuggingFace.

The source code to reproduce all plots in this manuscript is also available on Zenodo, and will see continuous updates on GitHub at \url{https://github.com/ddasilva/probabilistic-solar-wind}.

\section*{Conflict of Interest declaration}
The authors declare there are no conflicts of interest for this manuscript.


%
\bibliography{./references}
%


%
%
%
%
%

\end{document}